\documentclass[10pt, letterpaper]{article}

\let\n\noindent

\newcommand{\beq}{\begin{equation}} \newcommand{\eeq}{\end{equation}}
\newcommand{\beqa}{\begin{eqnarray}}    \newcommand{\eeqa}{\end{eqnarray}}
\newcommand{\btab}{\begin{tabular}}     \newcommand{\etab}{\end{tabular}}

\newcommand{\bt}{\begin{table}}     \newcommand{\et}{\end{table}}

\newcommand{\ba}{\begin{array}}     \newcommand{\ea}{\end{array}}
\newcommand{\bc}{\begin{center}}        \newcommand{\ec}{\end{center}}
\newcommand{\bfig}{\begin{figure}}      \newcommand{\efig}{\end{figure}}
\newcommand{\bp}{\begin{picture}}       \newcommand{\ep}{\end{picture}}
\newcommand{\bq}{\begin{quote}}     \newcommand{\eq}{\end{quote}}
\newcommand{\ben}{\begin{enumerate}}    \newcommand{\een}{\end{enumerate}}

\font\tenmsy=msbm10
\font\sevenmsy=msbm10 at 7pt
\font\fivemsy=msbm10 at 5pt
\newfam\msyfam 
\textfont\msyfam=\tenmsy
\scriptfont\msyfam=\sevenmsy
\scriptscriptfont\msyfam=\fivemsy
\def\blackB{\fam\msyfam\tenmsy}
\def\Z{{\blackB Z}}

\let\R\rangle

\def\frac#1#2{{\textstyle{#1\over #2}}}

\def\text#1{\quad\hbox{#1}\quad}

\def\e{\epsilon}
\def\ka{\kappa}

\def\At{{\tilde{A}}}

\def\Bt{{\tilde{B}}}

\def\ch{{\hat \chi}}

\def\y{{\infty}}

\def\rw{\rightarrow}

\def\R{\rangle}

\def\frac#1#2{{#1 \over #2}}

\def\ph{{\hat \phi}}
\def\ch{{\hat \chi}}

\def\SM{{\cal {SM}}}
\def\M{{\cal {M}}}
\def\uh{{\widehat u}}
\def\rw{{\rightarrow}}

\begin{document}

\vskip18pt

\title{\vskip60pt${\SM}(2,4\ka)$ fermionic characters and restricted jagged
partitions
    }
\vskip18pt

\author{J.-F. Fortin, P. Jacob and P.
Mathieu\thanks{jffortin@phy.ulaval.ca, pjacob@phy.ulaval.ca,
pmathieu@phy.ulaval.ca.  This work is supported by NSERC.} \\ 
\\
D\'epartement de physique, de g\'enie physique et d'optique,\\
Universit\'e Laval, \\
Qu\'ebec, Canada, G1K 7P4.
}

\date{June 2004}

\maketitle


\vskip0.3cm
\centerline{{\bf ABSTRACT}}
\vskip18pt
A derivation of the basis of states for the ${\SM}(2,4\ka)$
superconformal minimal models is presented. It relies on a general hypothesis
concerning the role of the null field of dimension $2\ka-1/2$. The basis
is expressed solely in terms of $G_r$ modes and it takes the form of
simple exclusion conditions (being thus a quasi-particle-type basis). Its
elements are in correspondence with
$(2\ka-1)$-restricted jagged partitions. The generating functions of the latter
provide novel fermionic forms for the characters of the irreducible
representations in both Ramond and Neveu-Schwarz sectors.

\n 


\newpage


\section{Introduction}

The non-unitary minimal models  ${\cal M} (2,2k+1)$ have a number of rather nice properties. Among them, we
single out the fact that the characters of each irreducible representation can be written in a product form\footnote{Those
 minimal characters that have a product form are investigated in \cite{CBF}.} as well as in a particularly
simple  positive multiple sum (i.e., fermionic form - cf. \cite{KKMM}). These two expressions (product and positive sum)
are nothing but the two sides of the Rogers-Ramanujan identities for
$k=2$ and their generalization, Gordon-Andrews identities, for
$k>2$ \cite{An,Andr} (see also \cite{KKMM}).  The combinatorial interpretation underlying the multiple sum
leads to a very simple description of the basis of states \cite{FNO}. The states in the
irreducible module associated to the primary field
$\phi_{1,s}$, with
$1\leq s\leq 2k$, are obtained by the application, on the highest-weight state $|\phi_{1,s}\R$,  of the ordered
sequence of Virasoro modes 
\beq \label{stri}
L_{-n_1}L_{-n_2}\cdots L_{-n_m}|\phi_{1,s}\R\; , \qquad n_1\geq n_2\geq \cdots n_m\geq 1\; ,
\eeq
with the addition of a generic constraint:
\beq\label{exclu}
n_i\geq n_{i+k-1}+2\; ,\eeq
together with a boundary condition, $n_{m-s+1}\geq 2$,  that singularizes each module.
The complete set of states is obtained by summing over $m$.
The boundary condition amounts to a constraint on the number of 1 at the rightmost end of the partition
$(n_1,\cdots, n_m)$ which is associated to the state (\ref{stri}), that is, a limitation on the number of $L_{-1}$
acting on the highest-weight state. It reflects 
 the presence of the generic 
singular vector (i.e., the one that appears for generic values of $p,p'$) at level
$s$ in the Verma module of $|\phi_{1,s}\R$. On the other hand, the main exclusion condition (\ref{exclu}) is linked to
the `equation of motion' of the model \cite{ZamoZ},
that is, the null field associated to the vacuum 
singular vector at level
$2k$, whose presence does depend upon the minimal character of the model, i.e., that $p$ and $p'$ be coprime integers
$\geq 2$. This `equation of motion' takes the form $(T^k)+\cdots= 0$, where $T$ is the energy-momentum tensor, its
$k$-th power being properly normal ordered. This condition induces a constraint on groups of $k$ contiguous Virasoro
modes in the `bulk' of sequences of modes of the form (\ref{stri}) \cite{FNO}. For
such groups, one state has to be removed at each level. For consistency, the
states that are to be excluded are sequences with indices `as equal as
possible'. For instance, if
$k=3$, we need to take out all sequences of states associated to partitions containing any of these three substrings:
\beq
(\cdots, n,n,n,\cdots) \;,\quad
(\cdots, n+1,n,n,\cdots) \;,\quad
(\cdots, n+1,n+1,n,\cdots) \;.\eeq
This, of course, is equivalent to imposing (\ref{exclu}) for $k=3$. 
The remarkable aspect of the ${\cal M}(2,2k+1)$ models is that the structure of all the irreducible modules is
completely determined by this sole condition, up to the boundary constraint \cite{FNO}. The sufficiency of
this exclusion constraint is a priori quite unexpected.

Notice, {\it en passant}, that the core condition (\ref{exclu}) also controls the structure of the fermionic form of the
$\Z_k$ parafermionic characters (which is the Lepowsky-Primc formula) \cite{LP} (see also \cite{JMb}).

Given that the $\SM(2,4\ka)$ superconformal minimal models are in many ways the natural generalization of the
$\M(2,2k+1)$ models, it is quite natural to see whether their characters would also be linked with a sort of
`fermionic' (i.e., $\Z_2$ graded) version of the Andrews-Gordon identities. Actually, results in that direction have
already been obtained. In \cite{Mel}, it has been pointed out that all $\SM(2,4\ka)$ characters do have
a product form.  Recall that  the superconformal minimal models are
labeled by the integers $p',\, p$,  with  $(p-p')/2$ and $p'$ relatively coprime and that the bosonic form of the
normalized  characters (i.e., without the $q^{h-c/24}$ prefactor)  is given by
\begin{equation}
\ch_{r,s}^{(p',p)}(q)= {(-q^{1-\e})_\y\over (q)_\y)}\sum_{n\in \Z}\left(q^{n(npp'+rp-sp')/2}-q^{(np'+r)(np+s)/2}
\right)\;,
\end{equation}
with $1\leq r\leq p'-1$ and  $1\leq s\leq p-1$. The parameter $\e$ is defined by
\begin{equation}
\e= \left\{\matrix{1/2 &{\rm (NS)}& r+s \;{\rm even} \cr 0&{\rm (R)}& r+s \;{\rm odd}\;, \cr}\right.\end{equation}
and the following notation has been used
\beq
 (a)_m\equiv (a;q)_m= \prod_{j=1}^m(1-aq^{j-1})\;.
 \eeq
Consider the case where $p'=2,\, p= 4\ka$. Using the Jacobi triple-product identity (e.g., \cite{Andr} Theo. 2.8):
\beq \label{Jac}
\sum_{n=-\y}^\y (-1)^n q^{n(n+1)/2}z^n= (qz)_\y\, (z^{-1})_\y\,(q)_\y\;,
\eeq the
above characters  can be written in  product form as \cite{Mel}
\begin{eqnarray}\label{recuA} 
\ch_{1,s}^{(2,4\ka)}(q)&=&\prod_{n\not = 2\;{\rm mod}\; 4\atop n\not = 0, \pm s\;{\rm mod}\;
4\kappa}^\y {1\over (1-q^{n/2})} \qquad ( s\, {\rm odd}\, \leq 2\ka -1)\cr
&=& \prod_{n=1}^\infty  (1+ q^n)  \prod_{n\not
= 0,
\pm s/2
\;{\rm mod}\, (2\ka)}^\infty  {1\over (1- q^n)} \qquad ( s\, {\rm even}\, \leq 2\ka-2 )\cr
 &=&
 \prod_{n\not = 0\;{\rm mod}\;
\kappa}^\infty {(1+ q^n) \over (1- q^n)} \qquad\qquad (s= 2\kappa)\;.
\end{eqnarray}

Moreover, this
product formulation  points toward the proper fermionic form of the character \cite{Mel}. (In the
NS sector for instance, this follows from the analytic G\"ollnitz-Gordon
theorem (see
\cite{Andr} sect 7.4).)
Again, the combinatorial interpretation underlying the multiple-sum (cf. \cite{Andr} Theo. 7.11) can be lifted to a
description of the basis of states \cite{Mel}.\footnote{We
stress that this is always the combinatorial interpretation of the positive sum, and not the one of the
product expression, that yields a generic exclusion condition (i.e., depending only upon $\ka$) holding in the bulk of
all irreducible modules.} The one that emerges here involves both the Virasoro modes and their fermionic partners. More
precisely, the various states in the NS irreducible $|\ph_{1,s}\R$ module are the partitions of
$n/2$ into ordered  positive parts
$r_i\in\Z_+/2$, with $r_i\geq r_{i+1}$, where no half-integer part is repeated and 
\beq\label{baseM}
r_j\geq r_{j+\ka-1}+1 \quad {\rm if}\quad r_j\in \Z_++\frac12\qquad {\rm or}\qquad r_j> r_{j+\ka-1}+1 \quad {\rm if}\quad
r_j\in
\Z_+\;.
\eeq
This is supplemented by  a boundary condition which is that there must be at most $i-1$ parts $\leq 1$ (which is thus a
constraint on the maximal numbers of $G_{-1/2}$ and $L_{-1}$ factors acting directly on the highest-weight state).
In the R sector, fermionic sums have  been conjectured for $s=2 $ and $ 2\ka$ in \cite{Mel}. A complete set of
expressions in the R sector has subsequently been displayed and proved in \cite{BMO,BM}. The construction of the fermionic
characters there relies on a counting problem inspired by the thermodynamics of the XXZ chain of spin 1. A different
basis of states, pertaining to both sectors,  is also presented in
\cite{BMO} (cf. eqs (7.6)-(7.7)), here again extracted from the underlying combinatorics.

In the present work, we reconsider the construction of the fermionic characters of the $\SM(2,4\ka)$ models 
from a completely different perspective. In contradistinction with the previous works
\cite{Mel,BMO,BM} where the basis is read off the positive-sum expressions, we derive (granting a general hypothesis) 
the basis and deduce from it the fermionic characters. This work fits within a  general program of trying to construct
 quasi-particle-type bases  of states from intrinsic conformal field theoretical considerations.

The first step is the derivation of a free basis (i.e., pertaining to the
$\ka\rw\y$ limit or a non-minimal model). The resulting basis turns out to be solely formulated in terms of the
$G_r$ modes. But this is not a standard ordered basis, one that could be associated to partitions. It is rather
expressed in terms of  weakly ordered sequences of non-negative integers (in the R sector) or half-integers (in the NS
sector). Precisely this kind of basis has already been encountered in our previous analysis of the graded parafermionic
models (associated to the coset
$\widehat{osp}(1,2)_k/\uh(1)$)
\cite{JM,BFJM}. These weakly ordered sequences have been dubbed {\it jagged partitions} (see also \cite{FJM}). 

Now, by analogy with the $\M(2,2k+1)$ case, we expect that the `super equation of motion' of the $\SM(2,4\ka)$
model, that is, the field version of singular vector at level $2\ka-1/2$ in the vacuum module
$|\ph_{1,1}\R$\footnote{Recall that  the reducible module associated to the field
 $\ph_{r,s}$ has two primary singular vectors at levels
$rs/2$ and
$(p'-r)(p-s)/2$.}
to determine completely the bulk structure of the irreducible modules. This is our working hypothesis. 
This null field enforces constraints on contiguous sequences of
$(2\ka-1)$
$G_r$ modes. It thus leads to restrictions on our free jagged-type basis. Again, one state  has to be excluded at each
level for sequences of $(2\ka-1)$
$G_r$ modes. Similar constraints have already been observed in
\cite{JM} (but for sequences of
$2k$ graded parafermionic modes). Since the restrictions described in \cite{JM} embody a consistent way of removing
one state per level for such weakly ordered partitions,  we then naturally guess that the basis is generated by the 
$K$-restricted jagged partitions
\cite{FJMa} for the case
$K=2\ka-1$ (instead of $2k$ as in \cite{JM}).  The generating functions for these restricted jagged partitions (for
both parities of
$K$) have already been constructed in
\cite{FJMa}. Up to some simple modifications, these become natural candidate
for the characters of the  $\SM(2,4\ka)$ irreducible modules. In the R sector, this resulting multiple sum is already
known to be expressible in a product form that is precisely the one in (\ref{recuA}) \cite{FJMa}. In the NS sector, we
end up with  multiple-sum expressions whose equivalence with the product (\ref{recuA}) is demonstrated. In both sectors, the
resulting fermionic characters are new.


\section{Vacuum irreducible module and jagged partitions}

Consider first the vacuum character in a generic superconformal model for which  $p$ and $p'$ are arbitrary. The
vacuum irreducible module
$|\ph_{1,1}\R=|0\R$ is obtained by factoring out the singular vector at level $1/2$, namely  $G_{-1/2}|0\R$.
The character is thus
\beq
\ch_{1,1}^\y(q)={(-q^{1/2})_\y\over (q)_\y} (1-q^{1/2})= {(-q^{3/2})_\y\over (q^2)_\y}\;,\eeq
(where the upper index ${\y}$ reminds of the generic nature of the module or, equivalently, the situation in which 
all but the first singular vector are pushed to infinity). The underlying basis of states can be derived by simple
manipulations based on the commutation relations (mimicking the
discussion of section (3.2) of \cite{JM}); it is spanned by the states
\beq \label{sequ}
G_{-{r_1}}G_{-{r_2}}\cdots G_{-{r_{m-1}}}G_{-{r_m}}|0\R\;,\eeq
with $r_j\in\Z+1/2$ and
\begin{equation}\label{oro}
 r_j\geq r_{j+1}-1\;,\qquad  \qquad r_j\geq r_{j+2}\;, \qquad\qquad   r_m\geq 3/2Ê\;,
\end{equation} 
the complete set of states being obtained by summing over $m$.
In other words, the sequences $(r_1,\cdots, r_m)$ associated to the states (\ref{sequ}) are obtained
 by adding ordinary
partitions (ordered sequences of positive integers) on the ground state
$(\cdots 3/2,1/2,3/2,1/2,3/2)$.\footnote{The necessity of weakly ordered conditions (cf. (\ref{oro})) in a basis formulated in
terms of
$G$ modes only can be seen rather simply form the first two non-zero states. The first non-vanishing state in the vacuum
module is
$G_{-3/2}|0\R$. The next one has to be $L_{-2}|0\R$. This can be rewritten solely in terms of $G$ modes quite simply  as
$G_{-1/2}G_{-3/2}|0\R$.}

 The sequence $(r_1,\cdots, r_m)$ looks like a
jagged partitions but with half-integers. To make the link precise, we subtract $1/2$ from each entries; the
ground state is then
$(\cdots 010101)$. With $r_i-1/2=n_i$, the conditions (\ref{oro}) become simply:
\begin{equation}\label{oror}
 n_j\geq n_{j+1}-1\;,\qquad  \qquad  n_j\geq n_{j+2}\;, \qquad\qquad   n_m\geq 1\;.
\end{equation}
But these are precisely the defining conditions for jagged partitions.\footnote{As it is clear from the ground-state
example, jagged partitions may contain some 0's, but since the rightmost entry $n_m$ has to be $\geq
1$, there can be no two adjacent 0's since $n_{m-2}\geq n_m$.} The generating functions for jagged
partitions of length $m$ and  weight $n=\sum_{j=1}^m n_j$, enumerated by $j(n,m)$, is \cite{FJM} (Theo.
16):
\beq
J(z;q)= \sum_{n,m\geq 0} j(n,m)z^mq^n= {(-zq)_\y\over (z^2q)_\y}\; .\eeq
That there is an underlying  fermionic structure is indicated by the numerator $(-zq)_\y$ which is the
generating function for partitions into distinct parts.

To go from this generating function $J(z;q)$ to the $\ch_{1,1}^\y$ character, one simply has to find the proper way of
modifying $J(z;q)$ to account for the addition of $1/2$ to each parts of the summed jagged partitions.  We notice
that we have the extra variable $z$ at our disposal, whose exponent keeps track of the length. Therefore, by
replacing $z^m$ by $q^{m/2}$, we perform precisely the required transformation.  We thus simply evaluate $J(z;q)$ at
$z=q^{1/2}$. And this indeed reproduces the above vacuum character:
\beq
 J(q^{1/2};q)= \ch_{1,1}^\y(q) \; .\eeq
This identity confirms the jagged-type nature of the generic vacuum-module basis.

Now let us turn our attention to the minimal models $\SM(2,4\ka)$ and see heuristically how the constraint on modes
at distance $2\ka-2$ pops up from our hypothesis on the role of the vacuum null field of dimension $2\ka-1$. The 
verification of this hypothesis will be established a posteriori, from the correctness of the resulting character. 


To see how the null field constraint can be implemented  in terms of an exclusion,
let us take for simplicity  the case where $\ka=2$. The vacuum null field is of dimension $7/2$ and it is of the form
$(TG)+\cdots =0$. But since our basis does not incorporate Virasoro modes,
$T$ must be reexpressed as a bilinear in $G$. This leads to a constraint on products of three adjacent $G$
factors. Granting that this `equation of motion' is responsible for all the constraints in the bulk, we see that in
the present case it leads  to restrictions on products of three modes, i.e., one state has to be
removed at each level in sequence of three $G_r$ modes.
The proper way of implementing this restriction
 is to throw away all sequences $(r_1,\cdots , r_m)$ containing any
of the three substrings:
\beq
(\cdots, r,r-1,r\cdots);\,\qquad (\cdots, r,r,r\cdots);\,\qquad (\cdots, r,r+1,r\cdots)\, .\eeq
This is a sort of generalized exclusion condition \cite{JM}. 

For a general value of $\ka$,
the generating singular vector at level $2\ka-1/2$ is associated to the null field $(T^{\ka-1}G)+\cdots=0$, which (by
resolving $T$ into a pair of $G$ factors) leads to constraints on contiguous $(2\ka-1)$ modes $G_r$.\footnote{As shown in
the appendix, the same argument applied to the free basis formulated in terms of $T$ and $G$ modes leads to the basis
(\ref{baseM}).} This restriction is taken into account by 
$(2\ka-1)$-restricted jagged partitions \cite{JM,BFJM,FJMa}.  These are defined as jagged partitions further subject to 
\begin{equation} \label{allo}
n_j \geq  n_{j+2\ka-2} +1 \qquad{\rm or} \qquad
 n_j = n_{j+1}-1 =  n_{j+2\ka-3}+1= n_{j+2\ka-2} \;.
\end{equation}The first condition imposes a difference 1 at distance $2\ka-2$. However, it is not exclusive due to the
presence of the second condition. The latter can also be written as follows: if $n_j = 
n_{j+2\ka-2}$, then
$n_{j+1}
\geq  n_{j+2\ka-3}+2$. In other words, if we have no difference at distance $2\ka-2$, we allow for an in-between difference
of 2 at distance $2\ka-4$. It is simple to see that the inequality in $n_{j+1} \geq  n_{j+2\ka-3}+2$ must in fact be an
equality. The allowed situation corresponds to the following case: $(\cdots, n,n+1,\cdots ,n-1,n,\cdots)$. For instance,
for $\ka=3$, so that $2\ka-1=5$, the above conditions amount to exclude all jagged partitions containing any of
\beq
(n,n-1,n,n-1,n)\,,\,(n,n,n,n-1,n)\,,\,(n,n,n,n,n)\,,\,(n,n+1,n,n,n)\,,\,(n,n+1,n,n+1,n)\; .
\eeq
For instance $(43232312)$ is $5$-excluded because $n_2=n_6$ but $n_3-n_5\not=2$.\footnote{Other illustrative examples can be
found in \cite{JM,BFJM,FJM,FJMa}.}

The generating function for $(2\ka-1)$ restricted jagged partitions has been obtained in \cite{FJMa} and it reads:
\beq
J_{2\ka-1}(z;q)= \sum_{m_1,\cdots,m_{\ka-1}=0}^\y {(-zq^{1+m_{\ka-1}})_\y\, 
q^{N_1^2+\cdots+ N_{\ka-1}^2}\; z^{2N} \over (q)_{m_1}\cdots (q)_{m_{\ka-1}} }\;,
\eeq
where $N_j$ and $N$ are given 
by
\begin{equation}\label{defNL}
N_j= m_j+\cdots
+m_{\ka-1} \,,  \qquad N=N_1+N_2+\cdots +N_{\ka-1} \;.
\end{equation}
To recover the vacuum character, we need to add up $1/2$ to each part of the summed jagged partitions, which
again amounts to set
$z=q^{1/2}$:
\begin{eqnarray}
J_{2\ka-1}(q^{1/2};q)&=&\sum_{m_1,\cdots,m_{\ka-1}=0}^\y {(-q^{3/2+m_{\ka-1}})_\y\, 
q^{ N+ N_1^2+\cdots+ N_{\ka-1}^2}\; \over (q)_{m_q}\cdots (q)_{m_{\ka-1}} }\cr
&=&\sum_{m_0,\cdots,m_{\ka-1}=0}^\y {
q^{m_0(m_0+2)/2+ 
 m_0m_{\ka-1}+ N+ N_1^2+\cdots+ N_{\ka-1}^2}\; \over (q)_{m_0}\cdots (q)_{m_{\ka-1}} }\;.
\end{eqnarray}
The second expression is obtained by means of the Euler relation (\cite{Andr}, Corr. 2.2)
\beq
(-t)_\y= \sum_{n\geq 0} {t^nq^{n(n-1)/2}\over (q)_n}\; .\eeq
Now, the following identity is demonstrated  in \cite{Bre} (eq 3.7) 
\begin{equation}\label{bres}
\sum_{m_1,\cdots,m_{\ka-1}=0}^\y {(-q^{3/2+m_{\ka-1}})_{\y}
q^{N +N_1^2+N_2^2+\cdots+ N_{\ka-1}^2}\;  \over (q)_{m_1}\cdots (q)_{m_{\ka-1}} }= 
\prod_{n\not = 2\;{\rm
mod}\; 4\atop n\not = 0, \pm 1\;{\rm mod}\; 4\kappa}^\y {1\over (1-q^{n/2}) }\;.
\end{equation}
Comparing the product on the rhs with the one appearing in  the first line (\ref{recuA}), we see that it corresponds to the
vacuum character of the
$\SM(2,4\ka)$ superconformal minimal model. We have thus established the following relation.
\beq
J_{2\ka-1}(q^{1/2};q)= \ch^{(2,4\ka)}_{1,1}(q) \;.\eeq
This demonstrates the correctness of our hypothesis, at least  for the basis of states in the vacuum module. Let
us now turn to the consideration of the other irreducible modules.

\section{$\SM(2,4\ka)$  irreducible modules and restricted jagged partitions}

Guided by the form of the $\M(2,2k+1)$ characters and our analysis of the $\SM(2,4\ka)$ vacuum module,  we expect that
 every $\SM(2,4\ka)$ irreducible module
will be expressed in terms of  $(2\ka-1)$-restricted jagged partitions (shifted by $1/2$ in the NS sector). Moreover, 
every module
$|\ph_{1,s}\R$  should be also characterized  by a boundary condition that reflects the existence of the (generic)
singular vector at level $s/2$. In the NS sector, this is a constraint of the number of $G_{-1/2}$ terms at the
rightmost end of the string; there must be at most $s-1$ of them. In the R sector, this is a  constraint of the
number of
$G_{0}G_{-1}$ pairs, which should not exceed $s/2-1$. 

The related generating functions have  been obtained previously \cite{FJMa}.
Let 
 $A_{2\ka-1,2i}(m,n)$ stands for the number of jagged partitions of length $m$ and
weight $n=\sum_{j=1}^m n_j$ satisfying 
the restrictions 
(\ref{allo}) and containing at most $i-1$ pairs 01. 
Its generating function is \cite{FJMa}
\begin{eqnarray}
\label{Amusi}
\At_{2\ka-1,2i}(z;q) &=& \sum_{n,m\geq 0} A_{2\ka-1,2i}(m,n) z^m q^n \cr
&=&   \sum_{m_1,\cdots,m_{\ka-1}=0}^\y {(-zq^{1+m_{\ka-1}})_\y
q^{ N_1^2+\cdots+ N_{\ka-1}^2+L_{i}}\; z^{2N} \over (q)_{m_1}\cdots (q)_{m_{\ka-1}} }\;,
\end{eqnarray}
where $N_j$ and $N$ are given 
by (\ref{defNL}) and $L_j$ is  
\begin{equation}\label{defL}
 L_j=N_j+\cdots +N_{\ka-1} \, \qquad \;  (L_\ka=L_{\ka+1}=0).
\end{equation}

According to the above discussion, we expect 
that in the R sector, the basis of
states to be simply the
$2\ka-1$ restricted jagged partitions with at most  $i-1$ of pairs 01 at the rightmost
end, with $i=s/2\leq \ka$. This translates into the following identification
\begin{equation}\label{Rca}
\At_{2\ka-1,2i}(1;q) = \ch_{1,2i}^{(2,4\ka)}(q) \;.\end{equation}
To check the  correspondence (\ref{Rca}), we use the reexpression of this multiple sum in terms of a product that
has been derived in \cite{FJMa,Bre,Bres}
\begin{eqnarray} 
\label{theor}
{\tilde A}_{2\ka-1,2i}(1;q)&=&  \prod_{n=1}^\infty  (1+ q^n)  \prod_{n\not
= 0,
\pm i
\;{\rm mod}\, (2\ka)}^\infty  (1- q^n)^{-1} \qquad \qquad ( i<\ka )\cr
{\tilde A}_{2\ka-1,2\ka}(1;q)&=& 
 \prod_{n\not = 0\;{\rm mod}\;
\kappa}^\infty {(1+ q^n)\over  (1- q^n)} \qquad \qquad 
\end{eqnarray}
The comparison  with (\ref{recuA}) implies the identification (\ref{Rca}).  Note that by using $(-q^{1+m})_\y=
(-q)_\y/(-q)_m$, we can rewrite $\At_{2\ka-1,2i}(1;q)$, or $\ch_{1,2i}^{(2,4\ka)}(q)$, as
\beq
\label{Amusie}
\ch_{1,2i}^{(2,4\ka)}(q) = (-q)_\y \sum_{m_1,\cdots,m_{\ka-1}=0}^\y {
q^{ N_1^2+\cdots+ N_{\ka-1}^2+L_{i}} \over (q)_{m_1}\cdots (q)_{m_{\ka-2}}(q^2;q^2)_{m_{\ka-1}} }\;.
\eeq
This expression for the $\SM(2,4\ka)$  R characters is different from the one presented in
\cite{Mel} (for $i=1,\, \ka$) and in \cite{BMO,BM} (for all $i$).  

Let us illustrate our interpretation of the R characters with a simple example. The two  R characters in the $\SM(2,8)$
model are 
\begin{eqnarray} \label{R28}
\ch_{1,2}^{(2,8)}(q)&=&\prod_{n\not= 0 \;{\rm mod 4}}^\infty {1\over  (1- q^n)}=
1+q+2q^2+3q^3+4q^4+6q^5+9q^6+12q^7+16q^8+\cdots\cr
\ch_{1,4}^{(2,8)}(q)&=& \prod_{n\; {\rm odd}\, \geq 1\;}^\infty {(1+ q^n)\over  (1- q^n)} = 1 + 2 q + 2 q^2  + 4
q^3  + 6 q^4  + 8 q^5  + 12 q^6  + 16 q^7  + 22 q^8  
+ \cdots
\end{eqnarray}
Consider the coefficient of $q^6$ in each case. In $\ch_{1,4}^{(2,8)}$, it counts the number $3$-restricted jagged partitions
containing at most one pair of 01 (with no length constraint). These are
indeed 12 of them:
\beq
\{(3201)\; (2301)\; (501)\; (411)\; (321)\; (231)\; (312)\; (51)\; (42)\; (33)\; (6)\}\;.
\eeq
The corresponding coefficient in $\ch_{1,2}^{(2,8)}$ counts the number $3$-restricted jagged
partitions but with no pair of 01. This eliminates the first three jagged partitions of the previous set, reducing thus the
number to 9.

We now turn to the description of the NS sector. Every state should be  associated to a $(2\ka-1)$-restricted jagged
partition made of half-integers, with a tail containing at most $s-1$ parts equal to 1/2. It is thus convenient to add
1/2 to each part (instead of subtracting 1/2 as we did in the previous section for the vacuum module) and get integer
parts ending with at most
$s-1$ parts equal to 1 (and by construction, there can be no 0, that is, no pair of 01). 

The generating functions for such
boundaries have also been computed in
\cite{FJMa}.  Let $ B_{2\ka-1,j} (m,n)$ be the number  of $(2\ka-1)$-restricted jagged  partitions of $n$  into $m$ 
parts with  at mostÊ
$(j-1)$ consecutiveÊ$\, 1$ at the right end, with $1\leq j\leq 2\ka-1$.  
Its generating function is
\begin{eqnarray}\label{grande}
& & \Bt_{2\ka-1,2i}(z;q)=\sum_{m_1,\cdots,m_{\ka-1}=0}^\y { (-zq^{1+m_{\ka-1}})_\y\,
q^{N_1^2+\cdots+ N_{\ka-1}^2+L_{i}+N}\; z^{2N} \over (q)_{m_1}\cdots (q)_{m_{\ka-1}}
}\;,\,
\end{eqnarray}
and 
$\Bt_{2\ka-1,2i\pm1}(z;q)$ is obtained from (\ref{grande}) and (\ref{Amusi}) via
\beq\label{graB}
 \Bt_{2\ka-1, 2i\pm1 } (z;q)=\Bt_{2\ka-1, 2i } (z;q) \pm  (zq)^{2i-(1\mp1)/2} \, \At_{2\ka-1, 2\ka-2i}
(zq;q)\;.
\eeq
Notice that $\At_{2\ka-1,0}(z;q)= \Bt_{2\ka-1,0}(z;q)=0$. 

Given that in the NS sector
$s$ odd, we see that the function of interest is $ \Bt_{2\ka-1, s} (z;q)$ with $s=2i-1$ and $1\leq i\leq \ka$. Still we
need to fix $z$ as a function of $q$. Since here we have added 1/2 to each part of the sequences $(r_1,\cdots , r_m)$ to get
the jagged partitions, then in order to recover the NS characters for the $\Bt_{2\ka-1, s} (z;q)$ function, we need to
subtract
$1/2$ from each part of the summed jagged partitions. This amounts to fixing
$z=q^{-1/2}$. We thus arrive at the identification 
\beq\label{NScar}
\Bt_{2\ka-1, 2i-1} (q^{-1/2};q) = \ch_{1,2i-1}^{(2,4\ka)}(q) \;.\end{equation}
With
\beq \label{recuba}
\Bt_{2\ka-1, 2i-1}(q^{-1/2};q)=\Bt_{2\ka-1, 2i}(q^{-1/2};q)- q^{i-1/2} \At_{2\ka-1, 2\ka-2i}(q^{1/2};q)\;,
\eeq
the resulting multiple-sum expression of the NS characters reads
\beq \label{ourNS}
\ch_{1,2i-1}^{(2,4\ka)}(q)=
\sum_{m_1,\cdots,m_{\ka-1}=0}^\y { (-q^{1/2+m_{\ka-1}})_\y\,
q^{N_1^2+\cdots+ N_{\ka-1}^2+L_i} \over (q)_{m_1}\cdots (q)_{m_{\ka-1}} }\left[1-
{q^{i-1/2+N+L_{\ka-i}-L_i}\over (1+q^{1/2+m_{\ka-1}})} \right]\;.
\eeq
This again differs from the expression presented in \cite{Mel,BMO,BM}.

To demonstrate  the identification (\ref{NScar}) for $1\leq i\leq\ka$,\footnote{Note that (\ref{NScar}) is readily
verified for the vacuum character.  In that case,
there are no $1/2$ at the end and using eq. (\ref{graB}) for $i=0$ (and the $+$ sign), we have
$\Bt_{2\ka-1, 1 } (q^{-1/2};q)=Ê   \, \At_{2\ka-1, 2\ka} (q^{1/2};q)\;.$
The boundary condition in $\At_{2\ka-1, 2\ka} (q^{1/2};q)$ is superfluous, being already taken into account
by the restriction at distance $2\ka-2$ (which implies that there can be at most $\ka-1$ pairs of 01 at the right end).
Therefore, $\At_{2\ka-1, 2\ka} (q^{1/2};q)$ is the generating functions for all $(2\ka-1)$-restricted jagged
partitions, with each part being augmented by 1/2 (since $z=q^{1/2}$). It is thus  precisely the function we called
$J_{2\ka-1}(z;q)$ in the previous section, proven there to be equal to $\ch_{1,1}^{(2,4\ka)}(q)$.} we need to show
that $\Bt_{2\ka-1, 2i-1} (q^{-1/2};q)$  can be written in the product form that corresponds to the first line of
(\ref{recuA}). For this we use results of Bressoud \cite{Bres}, who introduces the function $C_{\kappa ,i}$
defined as (cf. his eq (2.1))
\begin{eqnarray}\label{Cdef}
C_{\kappa ,i}(b^{-1},c^{-1};a;q) &=& \frac{1}{(a)_\infty}\sum_{n\geq 0}
\frac{(a)_n(b)_n(c)_n(-1)^n(1-a^iq^{2ni})a^{\kappa n}(bc)^{-n}q^{[n^2(2\kappa -1)+n(3-2i)]/2}}{(q)_n(aq/b)_n(aq/c)_n}\cr
 &=& \sum_{m_1,\ldots,m_{\kappa -1}\geq 0}\frac{(aq/bc)_{m_{\kappa -1}}q^{N_1^2+\cdots+N_{\kappa
-1}^2-(N_1+\cdots+N_{i-1})}a^{N_1+\cdots+N_{\kappa -1}}}{(q)_{m_1}\cdots(q)_{m_{\kappa -1}}(aq/b)_{m_{\kappa
-1}}(aq/c)_{m_{\kappa -1}}}\;.
\end{eqnarray}
The second line is lemma 2 in \cite{Bres}.
From the latter expression, we see that our functions $\tilde{A}$ and $\tilde{B}$ can both be written in terms of $C$
as follows:
\begin{eqnarray}
\tilde{A}_{2\kappa -1,2i}(z;q) &=&(-zq)_\infty C_{\kappa ,i}(-(zq)^{-1},0;z^2q;q)\cr
\tilde{B}_{2\kappa -1,2i}(z;q) &=&(-zq)_\infty C_{\kappa ,i}(-(zq^2)^{-1},0;(zq)^2;q)\;,
\end{eqnarray}
and, using (\ref{recuba}),
\beq
\tilde{B}_{2\kappa -1,2i-1}(q^{-1/2};q) = (-q^{1/2})_\infty C_{\kappa ,i}(-q^{-3/2},0;q;q)-(-q^{3/2})_\infty
q^{i-1/2}C_{\kappa ,\kappa -i}(-q^{-3/2},0;q^2;q)\;.
\eeq
From the definition of $C$ (i.e., the first line of (\ref{Cdef})), we have
\begin{eqnarray}
C_{\kappa ,i}(-q^{-3/2},0;q;q) &=&\frac{1}{(1+q^{1/2})(q)_\infty}\sum_{n\geq
0}(-1)^n(1+q^{n+1/2})(1-q^{(2n+1)i})q^{\kappa n^2+(\kappa -i-1/2)n}\cr C_{\kappa ,\kappa -i}(-q^{-3/2},0;q^2;q)
&=&\frac{1}{(q)_\infty}\sum_{n\geq 0}(-1)^n(1-q^{n+1})(1-q^{2(n+1)(\kappa -i)})q^{\kappa n^2+(\kappa +i-1/2)n}\;.
\end{eqnarray}
$\tilde{B}_{2\kappa -1,2i-1}(q^{-1/2};q)$ can thus be expressed as
\begin{eqnarray}
\tilde{B}_{2\kappa -1,2i-1}(q^{-1/2};q) &=& \frac{(-q^{3/2})_\infty}{(q)_\infty}\sum_{n\geq 0}
\left[(-1)^n(1+q^{n+1/2})(1-q^{(2n+1)i})q^{\kappa n^2+(\kappa -i-1/2)n}\right.\cr
 && \left. -(-1)^n(1-q^{n+1})(1-q^{2(n+1)(\kappa -i)})q^{\kappa n^2+(\kappa +i-1/2)n+i-1/2}\right]\cr
 &=& \frac{(-q^{3/2})_\infty}{(q)_\infty}\sum_{n\geq 0}(-1)^nq^{\kappa n^2+(\kappa -i-1/2)n}\left[(1+q^{n+1/2})(1-q^{(2n+1)i})\right.\cr
&& \left.-(1-q^{n+1})(1-q^{2(n+1)(\kappa -i)})q^{(2n+1)i-1/2}\right]\; .
\end{eqnarray}
We then expand the square bracket, add $0=q^n-q^n$ and regroup terms in a suitable way:
\begin{eqnarray}
& & 1+q^{n+1/2}-q^{(2n+1)i}-q^{(2n+1)i-1/2}+q^{2\kappa (n+1)-i-1/2}-q^{2\kappa (n+1)+n-i+1/2}\cr
&=& (q^n+q^{n+1/2}-q^{(2n+1)i}
-q^{(2n+1)i-1/2})+(1-q^n)+(q^{2\kappa (n+1)-i-1/2}-q^{2\kappa (n+1)+n-i+1/2})\cr
&=& q^n(1+q^{1/2})(1-q^{(n+1/2)(2i-1)}+(1-q^n)+q^{2\kappa (n+1)-i-1/2}(1-q^{n+1})\;.
\end{eqnarray}
This allows us to write
\begin{eqnarray}
\tilde{B}_{2\kappa -1,2i-1}(q^{-1/2};q)
 &=& \frac{(-q^{1/2})_\infty}{(q)_\infty}\sum_{n\geq 0}(-1)^nq^{\kappa n^2+(\kappa -i+1/2)n}(1-q^{(n+1/2)(2i-1)})\cr
 && +\frac{(-q^{3/2})_\infty}{(q)_\infty}\sum_{n\geq 0}(-1)^nq^{\kappa n^2+(\kappa -i+1/2)n}(1-q^n)\cr
 && +\frac{(-q^{3/2})_\infty}{(q)_\infty}\sum_{n\geq 0}(-1)^nq^{\kappa (n+1)^2+(\kappa -i+1/2)(n+1)}(1-q^{n+1})\cr
 &=& \prod_{n\neq2\;{\rm mod}(4)\atop n\neq0,\pm(2i-1)\;{\rm mod}(4\kappa )}{1\over (1-q^{n/2})}\;.
\end{eqnarray}
We  use the Jacobi triple-product identity (\ref{Jac}) to transform the first sum and notice that  the other  two sums
cancel each other (i.e., letting 
$n\rightarrow n-1$ in the last one and noticing that for $n=0$, the first term of the second sum vanishes). This completes the proof of
(\ref{ourNS}). Note that according to the analytic version of the  G\"ollnitz-Gordon theorem (eq. (7.4.4) of \cite{Andr}), we have also
\beq \label{melz}
\prod_{n\not = 2\;{\rm mod}\; 4\atop n\not = 0, \pm s\;{\rm mod}\;
4\kappa}^\y {1\over (1-q^{n/2})}=  \sum_{m_1,\cdots,m_{\ka-1}=0}^\y {(-q^{1/2})_{N_1}
q^{ \frac12 N_1^2+N_2^2+\cdots+ N_{\ka-1}^2+L_{i}}\; \over (q)_{m_1}\cdots (q)_{m_{\ka-1}}
}\;.
\eeq
This last expression is the one appearing in \cite{Mel,BMO,BM}.

We end this section by illustrating our construction for the two NS characters of the $\SM(2,8)$ model:
\begin{eqnarray} \label{cads}
\ch_{1,1}^{(2,8)}(q)&=& \prod_{n\not
= 0\;\pm 1\; \pm 2\;{\rm mod}\, 8}^\infty  {1\over (1- q^{n/2})}\cr &=& 
1+q^{3/2}+q^2+ q^{5/2}+q^3+q^{7/2}+ 2q^4+2q^{9/2} +2q^5 +\cdots\cr
\ch_{1,3}^{(2,8)}(q)&=&  \prod_{n\not
= 0\;\pm 2\; \pm 3\;{\rm mod}\, 8}^\infty  {1\over (1- q^{n/2})}\cr &=&
1+q^{1/2}+q+q^{3/2}+2q^2+2q^{5/2}+2q^3+3q^{7/2}+4q^4+5q^{9/2}+5q^5\cdots
\end{eqnarray}
The coefficient of $q^n$ $(n>0)$ in $\ch_{1,3}^{(2,8)}$ counts the number of $3$-restricted jagged partitions made of
half-integers and containing at most two $1/2$ at the right end. The complete list of the contributing jagged partitions up
to level 5 is (with a  double bar $\|$  separating the contributions at different levels):
\begin{eqnarray}\label{exdeNS}
& & \{(1/2)\, \| \,
(1/2,1/2)\,\|\,(3/2)\,\|\,(1/2,3/2)\,(3/2,1/2)\,\|\,(3/2,1/2,1/2)\,(5/2)\,\|\cr &&\;\, (3/2,3/2)\,(5/2,1/2)\,\|\,
(5/2,1/2,1/2)\,(3/2,3/2,1/2)\,(7/2)\,\|\cr&&\;\,(3/2,3/2,1/2,1/2)\,(5/2,3/2)\,(3/2,5/2)\,(7/2,1/2)\,\|\cr
&& \;\,(7/2,1/2,1/2)\,(5/2,3/2,1/2)\,(5/2,1/2,3/2)\,(3/2,5/2,1/2)\,(9/2)\, \|\cr
&& \;\,(5/2,3/2,1/2,1/2)\,(3/2,5/2,1/2,1/2)\, (7/2,3/2)\,(5/2,5/2)\,(9/2,1/2)\,\}\;.
\end{eqnarray}
To recover the coefficients of $q^n$ in $\ch_{1,3}^{(2,8)}$, one simply counts  all sequences that do not contain 1/2 at the
end.

\section{Conclusion}

Inspired by the analysis of the $\M(2,2k+1)$ minimal models, where the
irreducible modules are described by a quasi-particle basis with an
exclusion condition at distance $k-1$ rooted in the presence of the null field
 at level
$2k$, we have obtained an analogous basis of states for the ${\SM}(2,4\ka)$
superconformal minimal models. The relevant null field in this context is the
one associated to the vacuum singular vector at level $2\ka-1/2$. It provides
constraints on a free basis of states described in terms of the $G_r$
modes only, a basis whose elements are in correspondence with jagged partitions
\cite{JM,BFJM,FJM,FJMa}.  The null field implies a restriction at distance
$2\ka-2$ on these partitions. States are then described by $(2\ka-1)$-restricted
jagged partitions. We recall that $2\ka$ restricted partitions have already
appeared in the description of the quasi-particle basis of $\Z_\ka$ graded
parafermionic models \cite{JM,BFJM}. It is thus quite interesting to see them
reappearing in a new context and with moduli of a different parity. 

Once the basis is fully characterized, the counting of states in irreducible
modules gets mapped into a combinatorial problem, that is, the
enumeration of restricted jagged partitions.  The next step amounts thus to
finding the corresponding generating functions. To tackle this problem, it is
standard (see e.g., \cite{Andr}, chap 7) to introduce partitions with particular
boundary conditions (i.e., number of 1 for  the usual partitions subject to
(\ref{exclu}) or the number of 01 or 1 for the restricted jagged partitions).
This provides a further degree of freedom that permits the derivation of
recurrence relations. These relations are central because the generating
functions are obtained through their solutions (cf. {\cite{BFJM,FJMa}). Thus,
the combinatorics lead naturally to generating functions with boundary terms.
Quite remarkably, these boundary conditions are precisely what distinguishe
the different irreducible modules.\footnote{This is also the case for the
$\M(2,2k+1)$ models
\cite{FNO}, the $\Z_k$ parafermionic models \cite{LP,JMb} and their graded
versions
\cite{JM}.} 

These generating functions appear in the form of positive multiple
sums. They provide thus fermionic-type characters for  ${\SM}(2,4\ka)$
superconformal irreducible modules. These formulas differ from those
previously presented \cite{Mel,BMO,BM}. They furnish thus new fermionic forms.

Finally, let us  stress that the fermionic characters displayed here results from a
derivation of the basis and that the latter is obtained  by conformal field
theoretical methods. In particular, it does not rely on the representation of
the conformal field theory as a solvable statistical model or an integrable
spin chain. 

\appendix

\section{The G\"ollnitz-Gordon-type NS basis}

Instead of describing the NS free basis in terms of the $G$ modes only, one could use the usual one defined by the modes
$L_r$ and
$G_r$, the latter being half-integers. This basis is generated by ordered sequences of mixed modes with no
half-integer mode being repeated. Consider then the effect of the null field $(T^{\ka-1}G)+\cdots = 0$ on this free basis.
It excludes groups of $\ka$ modes of the form $(n,\cdots,n ,n-1/2)$ and
$(n+1/2,n\cdots ,n)$, with $n$ integer. Acting with
$G_{-1/2}$ on the null field (roughly, transforming $G$ into $T$) leads to the exclusion of groups of $\ka$ terms of
the type
$(n,n,n\cdots ,n)$, $(n+1,n,n\cdots ,n)$, 
$(n+1,n+1,n\cdots ,n)$, etc. In the last sequences, there may be a mode $n+1/2$ in-between the $n+1$'s and $n$'s.
This clearly reproduces the basis (\ref{baseM}) of \cite{Mel}.
In a sense, this  basis is
the reformulation of ours in terms of overpartitions \cite{Love} (see in particular Sect. 6 of \cite{FJMa}).} On the other
hand, the basis introduced in \cite{BMO}, lifted from \cite{Burge}, does not seem to have a conformal field theoretical
interpretation.

\noindent {\bf ACKNOWLEDGMENTS}

We would like to thank to O. Warnaar for his useful comments on  the article.

\end{document}